\documentclass[fleqn,twoside]{article}
\usepackage{espcrc2}

\usepackage{graphicx}
\usepackage[figuresright]{rotating}
\usepackage{amsmath}

\title{The radial distributions of a heavy-light meson on a lattice}

\author{UKQCD Collaboration,
        J. Koponen\address[HY]{Department of Physical Sciences and
        Helsinki Institute of Physics, \\
        P.O. Box 64, FIN-00014 University of Helsinki, Finland},
        A. M. Green\addressmark[HY],
        C. Michael\address{Department of Mathematical Sciences, \\
        University of Liverpool, L69 3BX, UK}
        and
        P. Pennanen\addressmark[HY]
}

\begin{document}

\begin{abstract}
In an earlier work \cite{green:2001un},
the charge (vector) and matter (scalar) radial distributions of
heavy-light mesons were measured in the quenched approximation
on a $16^3\times 24$ lattice with $\beta=5.7$, a lattice spacing
of $a\approx 0.17$~fm, and a hopping parameter corresponding
to a light quark mass about that of the strange quark.

Several improvements are now made \cite{green:2002}:
\begin{enumerate}

\item
The configurations are generated using dynamical
fermions with quark-gluon coupling $\beta=5.2$
($a\approx 0.14$~fm);

\item
Many more gauge configurations are included (78
compared with the earlier 20);

\item
The distributions at many off-axis, in addition to on-axis,
points are measured;

\item
The data analysis is much more complete. In particular,
distributions involving excited states are extracted.
\end{enumerate}

The exponential decay of the charge and matter distributions can be
described by mesons of mass 0.9$\pm 0.1$ and 1.5$\pm 0.1$
GeV respectively --- values that are consistent with those
of vector and scalar $q\bar{q}$-states calculated {\em directly}
with the same lattice parameters.
\vspace{1pc}
\end{abstract}

\maketitle

\section{INTRODUCTION}

A knowledge of the radial distributions of quarks inside hadrons
could lead to a better understanding of the QCD description of
these hadrons and possibly suggest forms for phenomenological
models. As a step in this direction, we have measured the charge
(vector) and matter (scalar) densities of a heavy-light meson.

More explicitly, the heavy-light meson is simplified to being
an infinitely heavy quark ($Q$) and an antiquark ($\bar{q}$) with a
mass approximately equal to that of the strange quark. The physical
meson nearest to this idealised meson is the $B_{{\rm s}}$(5.37 GeV).
We consider two flavours of dynamical quarks ($u$ and $d$).

\section{MEASUREMENTS}

\subsection{Energies}

\begin{figure}
\centering
\includegraphics[width=0.99\columnwidth]{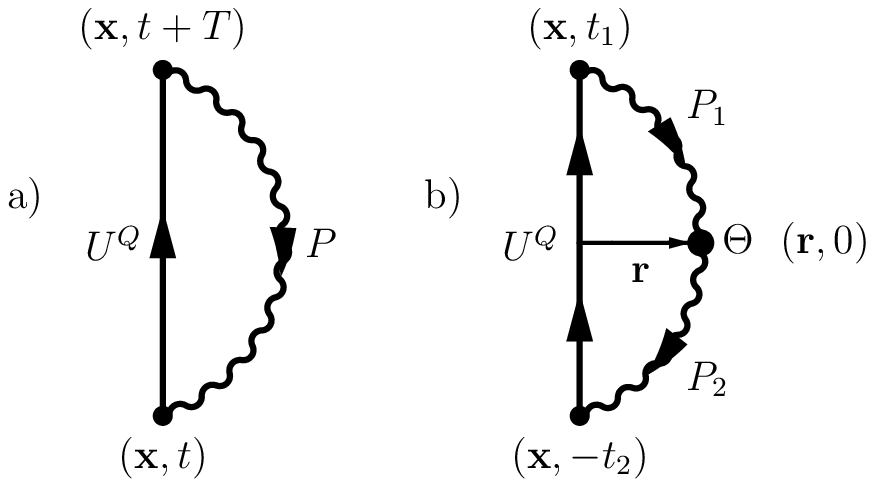}
\caption{Two- and three-point correlation functions
         $C_2(T)$ and $C_{3,c(m)}(T,\mathbf{r})$.}
\label{diagram_fig}
\end{figure}

The basic quantity for evaluating the energies of a heavy-light meson is
the 2-point correlation function -- see Fig.~\ref{diagram_fig} a) --
\begin{equation}
C_2(T)=\langle U^Q(\mathbf{x},t,T)P(\mathbf{x},t+T,t)\rangle,
\end{equation}
where $U^Q(\mathbf{x},t,T)$ is the heavy (infinite mass)-quark propagator
and $P(\mathbf{x},t+T,t)$ the light anti-quark propagator.
The $\langle ...\rangle$ means the average over the whole lattice.

\subsection{Heavy-light radial correlations}

When measuring the radial correlations, the basic quantity is the
3-point correlation function -- see Fig.~\ref{diagram_fig} b) --
\begin{equation}
C_{3,c(m)}(T, \mathbf{r})=\langle U^Q P_1\Theta P_2\rangle,
\end{equation}
where the $P_{1,2}$ are the light anti-quark propagators
that go from the $Q$ at time $t_1$ to the point $\mathbf{r}$
at $t=0$ and then return to $Q$ at time $-t_2$. Here $T=t_1+t_2$.
The probe $\Theta$ at $\mathbf{r}$ is here considered to
have two forms:
i) $\Theta=\gamma_4$ for measuring the
charge distribution (c) of the $\bar{q}$ and
ii) $\Theta=1$ for measuring the matter distribution (m).

The strategy is to first fit the $\langle C_2(T)\rangle$
by the approximate expression
\begin{equation}
\tilde{C}_2(T)=\sum_{\alpha}v^{\alpha}\mathrm{e}^{-E_{\alpha}T}v^{\alpha}.
\end{equation}
This results in the energy eigenvalues $E_{\alpha}$ and their
eigenvectors $v^{\alpha}$ (Ref.~\cite{MetP}).

Given the $E_{\alpha}$ and $v^{\alpha}$, the $C_{3,c(m)}(T, \mathbf{r})$ are then
fitted by
\begin{equation}
\tilde{C}_{3,c(m)}(T,\mathbf{r})=\sum_{\alpha\beta}v^{\alpha}\mathrm{e}^{-E_{\alpha}t_1}
x^{\alpha\beta}(r)\mathrm{e}^{-E_{\beta}t_2}v^{\beta},
\end{equation}
where the $x^{\alpha\beta}(r)$ are varied giving directly the desired
charge or matter density. In this work two levels of fuzzing with 2
and 8 iterations were used in addition to the unfuzzed lattice. This
enabled us to extract information of the excited states as well.

\section{RESULTS}

\subsection{Fits to the charge and matter densities}

It is convenient to parametrize the radial correlations
in some simple way. We have considered pure exponential
and Yukawa fits to the data as well as their lattice forms
\begin{equation}
\label{LE}
\frac{\pi a}{2r^{\mathrm{E}}L^3}
\sum_{\mathbf{q}}\frac{\cos(\mathbf{r}\cdot\mathbf{q})%
}{\bigl[\sum^3_{\substack{i=1}}\sin^2(\frac{aq_i}{2})+(\frac{a}{%
2r^{\mathrm{E}}})^2\bigr]^2}
\end{equation}
(lattice exponential) and
\begin{equation}
\label{LY}
\frac{\pi}{aL^3}
\sum_{\mathbf{q}}\frac{\cos(\mathbf{r}\cdot\mathbf{q})%
}{\sum^3_{i=1}\sin^2(\frac{aq_i}{2})+(\frac{a}{%
2r^{\mathrm{Y}}})^2}
\end{equation}
(lattice Yukawa).

\begin{figure}
\centering
\includegraphics[width=0.99\columnwidth]{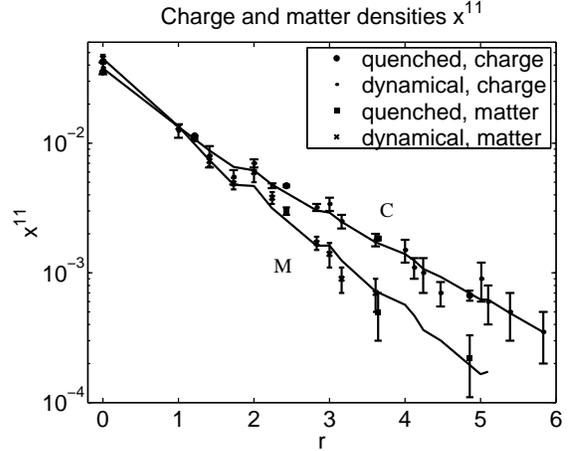}
\caption{The radial distribution of the ground
         state charge (C) and matter (M) densities.  The lattice
         exponential fits are plotted with solid lines. Here
         $r$ is in lattice units of $a\approx 0.14$~fm.}
\label{x11_fig}
\end{figure}

\begin{figure}
\centering
\includegraphics[width=0.99\columnwidth]{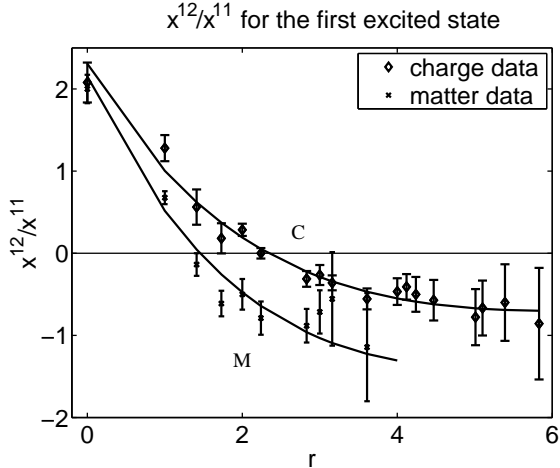}
\caption{The ratio $x^{12}/x^{11}$ for charge (C) and matter (M)
         densities. The solid curves guide the eye.}
\label{x12_fig}
\end{figure}

\begin{figure}
\centering
\includegraphics[width=0.99\columnwidth]{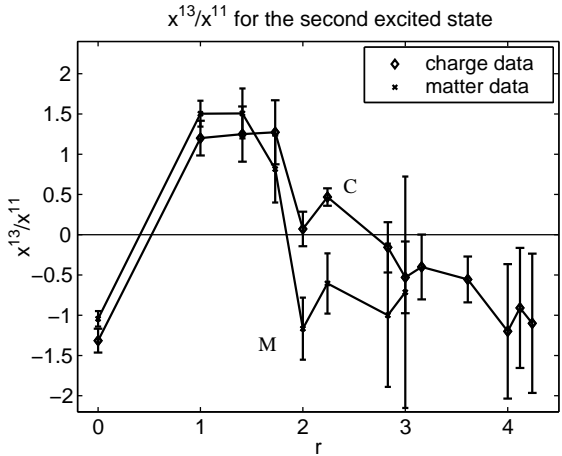}
\caption{The ratio $x^{13}/x^{11}$ for charge (C) and matter (M)
         densities.}
\label{x13_fig}
\end{figure}

They all fit the data well, and the lattice forms are able to
reproduce some of the structure (lattice artefacts) ---
see Fig.~\ref{x11_fig}.
The exponential decay of the charge and matter
distributions can be described by mesons of mass
$0.9\pm 0.1$ and $1.5\pm 0.1$~GeV respectively.
The matrix elements with excited states are shown
in Figs.~\ref{x12_fig},~\ref{x13_fig}.

\subsection{Sum rules}

It is also of interest to consider the sum rules that sum over all
values of $r$ i.e.
\begin{equation}
I_{\textrm{c(m)}}=\frac{\langle\sum_{\mathbf{r}}C_{3,c(m)}(T,\mathbf{r})
\rangle} {\langle C_2(T)\rangle}.
\end{equation}
For the charge and matter densities this yields $I_{\textrm{c}}=1.4(1)$ and
$I_{\textrm{m}}=0.9(1)$ respectively. The corresponding values for the
quenched approximation are $I_{\textrm{c}}=1.30(5)$ and $I_{\textrm{m}}=0.4(1)$.
By charge conservation we should get $I_{\textrm{c}}=1$ in the continuum limit.
Renormalisation effects of $\approx 0.7$ enter due to the finite
lattice spacing.

\section{CONCLUSIONS AND FUTURE}

The $S$-wave charge and matter densities can be measured quite
reliably out to $\approx 0.8$~fm and $\approx 0.5$~fm respectively.
No difference can be seen when comparing dynamical and quenched fermions.
This is probably due to the rather heavy mass of the light quarks.

This work is by no means the last word on the subject, and a lot remains
to be done. The next step could be, for example, measuring the $P$-, $D$-
and $F$-wave densities. The correlations in the baryonic and
$(Q^2\bar{q}^2)$ systems should be studied as well. We would also like
to understand the densities phenomenologically using the Dirac equation.
To check the continuum limit we should, of course, repeat the measurements
using larger $\beta$ and bigger lattices.

The authors wish to thank the Center for Scientific Computing in Espoo,
Finland for making available resources without which this project could
not have been carried out.
One of the authors (J.K.)
wishes to thank the Magnus Ehrnrooth Foundation
for financial support.

\end{document}